\documentclass{elsart1p}
\usepackage{graphicx}
\usepackage{amssymb}
\usepackage{amsmath}
\begin{document}
\begin{frontmatter}
%
%
%
%
%
\title{QCD phase diagram using PNJL model with eight-quark interactions}
%
%

\author{\bf{Paramita Deb}},
\ead{paramita.deb83@gmail.com}
\author{Abhijit Bhattacharyya}
\address{Department of Physics, University of Calcutta,
92, A. P. C. Road, Kolkata - 700009, INDIA}
\author{Sanjay K. Ghosh},
\author{Rajarshi Ray},
\author{Anirban Lahiri}
\address{Center for Astroparticle Physics \& 
Space Science, Bose Institute, Block-EN, Sector-V, Salt Lake, Kolkata-700091, INDIA}

\begin{abstract}
 We present the phase diagram and the fluctuations of different conserved 
charges like quark number, charge and strangeness at vanishing chemical 
potential for the 2+1 flavor Polyakov Loop extended Nambu--Jona-Lasinio model 
with eight-quark interaction terms using three-momentum cutoff regularisation.
The main effect of the higher order interaction term is to shift the critical 
end point to the lower value of the chemical potential and higher value of the 
temperature. The fluctuations show good qualitative agreement with the 
lattice data. 
 
\end{abstract}

\end{frontmatter}

\section{Introduction}
\vskip -10pt
Novel phases of strongly interacting matter at high temperature and density 
are very interesting area to study. The properties of the strongly interacting 
matter can be studied through effective models of QCD. Polyakov loop enhanced
Nambu--Jona-Lasinio model is one such model that successfully captures 
various properties of strongly interacting matter \cite{fuku,ratti,ray,deb1}. 
The 2+1 flavor PNJL and NJL model have been studied with 
four-quark and six-quark interaction terms in the Lagrangian \cite{deb1}. 
However the six-quark interaction imposes a serious problem of vacuum 
instability in the model which can be solved by introducing higher order 
eight-quark interaction terms in the Lagrangian \cite{osipov1,osipov2}. 
The search for the possible location of the critical end point (CEP) is one of 
the main issue of strong interaction physics. Also, fluctuations
and the correlations of conserved charges and their higher order cumulants
provide information about the degrees of freedom of strongly interacting 
matter. The study of diagonal and off-diagonal susceptibilities from the PNJL
model can also provide the information about the existence
of critical behavior \cite{deb3,deb4}. Here we give a detailed analysis of 
the CEP and the fluctuations of different conserved charges at zero chemical 
potential.

\begin{figure}
\centering
\includegraphics[scale=0.58,angle=270]{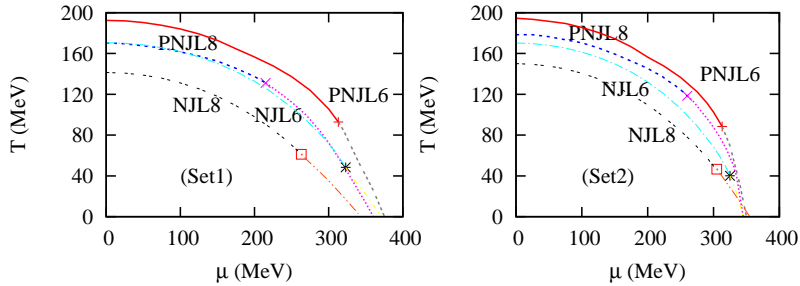}
\caption{Phase diagram in $\mu$ with $T$ for NJL and  PNJL
mode. }
\label {tmnew}
\end{figure}

{\section {Phase diagram}}
\vskip -10pt
The phase diagram and the location of the critical end point (CEP) are the
most important issue of strong interaction physics.
CEP is the point which separate the cross-over
transition from the first order phase transition. 
Here we study the phase diagram of strongly interacting matter using both 
NJL and PNJL model. We have used the nomenclature NJL6 and PNJL6 for the 
six-quark (6q) and NJL8 and PNJL8 for the eight-quark (8q) interactions. 
The details of the model and parameter set may be 
obtained in ref. \cite{deb2}. In fig. \ref{tmnew}
we have shown the phase diagrams of NJL and PNJL model with 6q and 8q 
type interactions. The parameter set are taken from \cite{deb2}. 
The introduction of the eight-quark 
interaction terms shift the CEP to lower chemical potential and the higher
temperature value. The recent lattice results show that the possible region 
of CEP should lie in $\mu_C/T_C \le 2.5$ \cite{ejiri}, which can only be
satisfied if the eight-quark interaction is added in the Lagrangian. The
values of the CEP for 
\begin{eqnarray}
(NJL6) (\mu_C,T_C)_{(set1)} \rm MeV =(323, 48.5), 
(\mu_C,T_C)_{(set2)} \rm MeV = (325, 40.15), \nonumber\\ 
(NJL8)  (\mu_C,T_C)_{(set1)} \rm MeV =(263, 61.2),  
(\mu_C,T_C)_{(set2)} \rm MeV =(305, 46.55), \nonumber\\ 
(PNJL6) (\mu_C,T_C)_{(set1)} \rm MeV =(313, 92.85), 
(\mu_C,T_C)_{(set2)} \rm MeV =(313, 88.5),\nonumber\\ 
(PNJL8)  (\mu_C,T_C)_{(set1)} \rm MeV = (260, 118.5),  
(\mu_C,T_C)_{(set1)} \rm MeV = (237, 122.05). \nonumber\\ 
\end{eqnarray}

{\section {Correlation of conserved charges and Specific heat}}
\vskip -10pt
The fluctuations and the correlation of conserved charges are sensitive
indicators of the transition from hadronic matter to the quark-gluon phase.
The definition of different cumulants  may be obtained in ref. \cite{deb3}. 
Here we will use the expansion around $\mu_X=0$, where the odd terms
vanish due to CP symmetry.
\begin{figure}[t]
\centering
\includegraphics[scale=0.58]{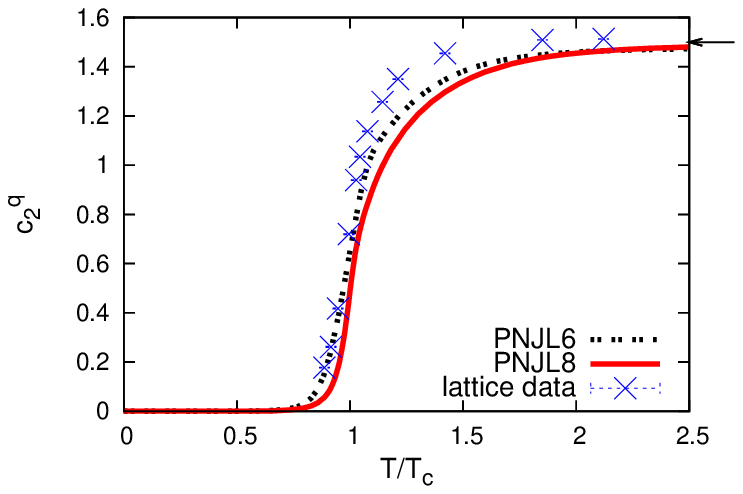}
\includegraphics[scale=0.58]{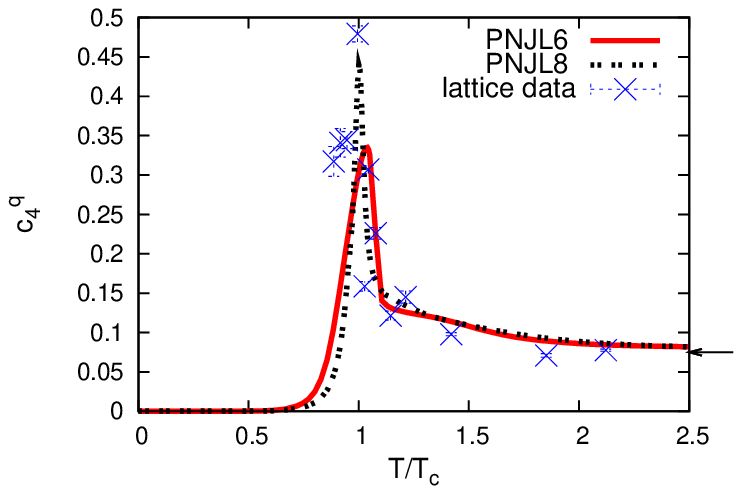}
\includegraphics[scale=0.58]{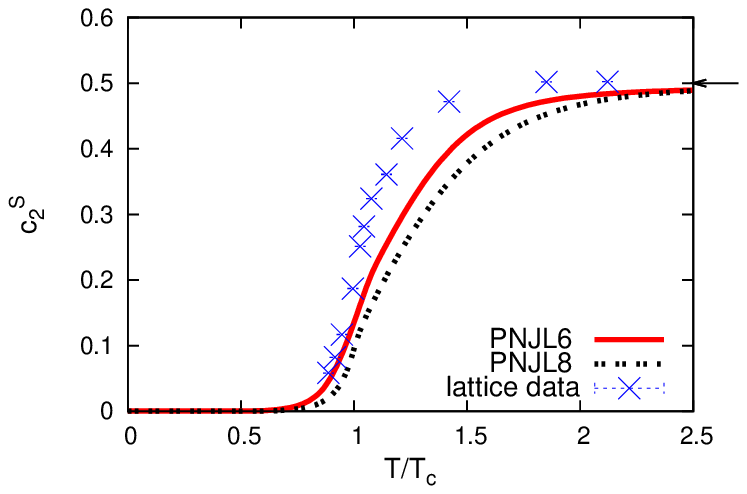}
\includegraphics[scale=0.58]{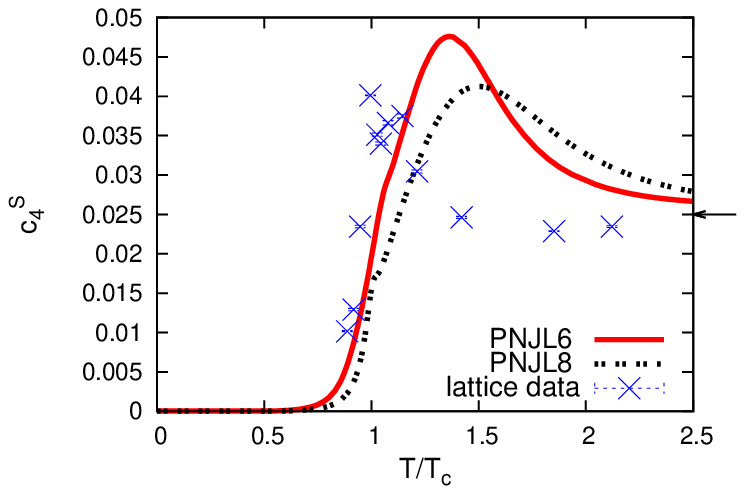}
\includegraphics[scale=0.58]{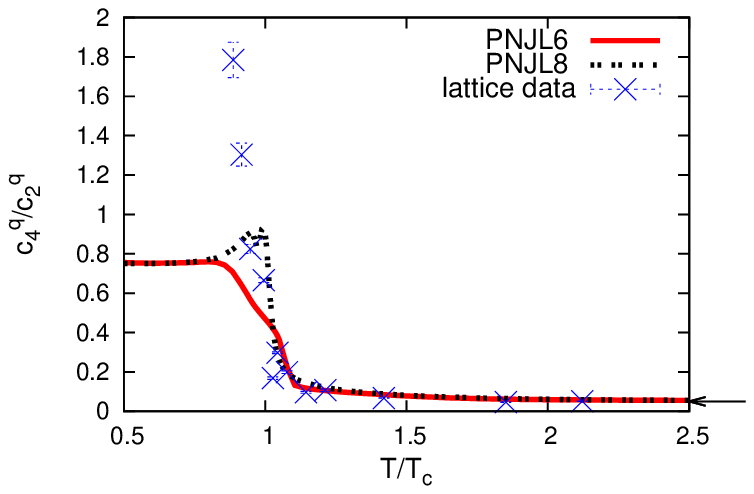}
\includegraphics[scale=0.58]{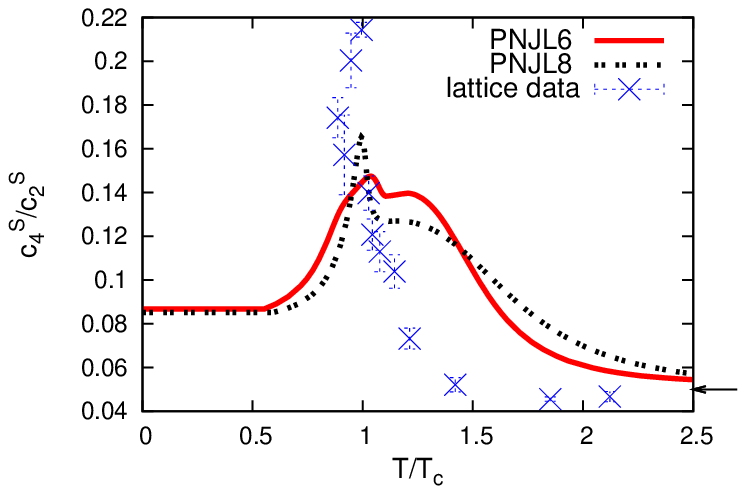}
\begin{minipage}{6in}
\caption{ {\small Variation of $c_2$, $c_4$ and Kurtosis 
with $T/T_C$, for
$\mu_X=\mu_q$, $\mu_S$ for 6q and 8q case. 
The lattice data taken from Ref. \cite{cheng}.}}
\label {qns}
\end{minipage}
\end{figure}
\begin{figure}[t]
\centering
\includegraphics[scale=0.58]{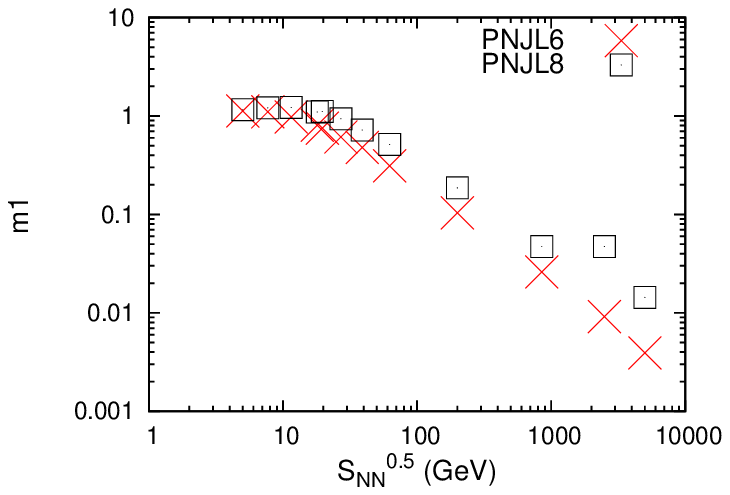}
\includegraphics[scale=0.58]{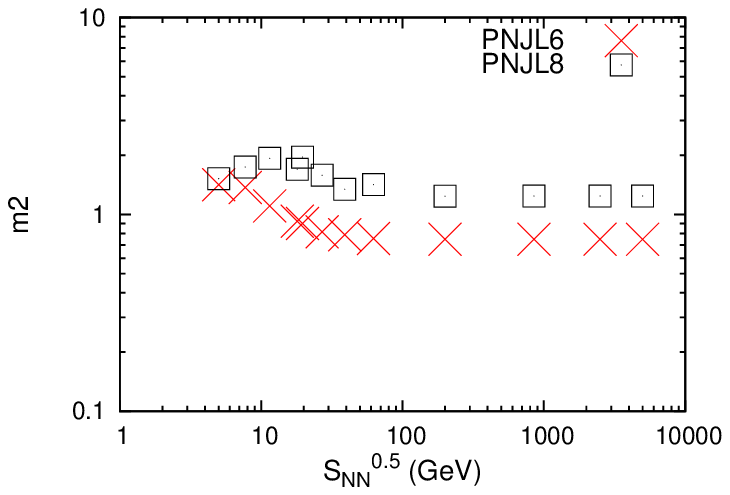}
\includegraphics[scale=0.58]{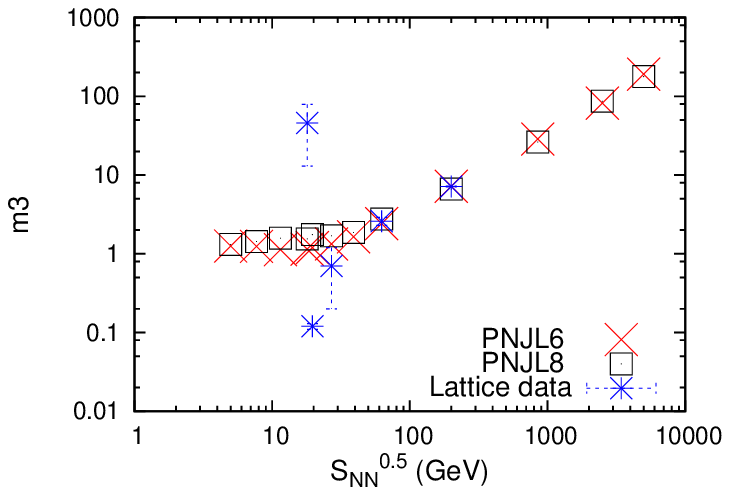}
\includegraphics[scale=0.58]{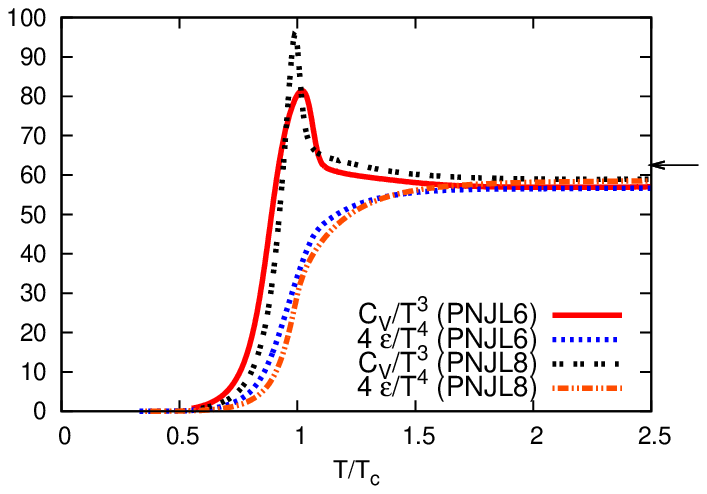}
\includegraphics[scale=0.58]{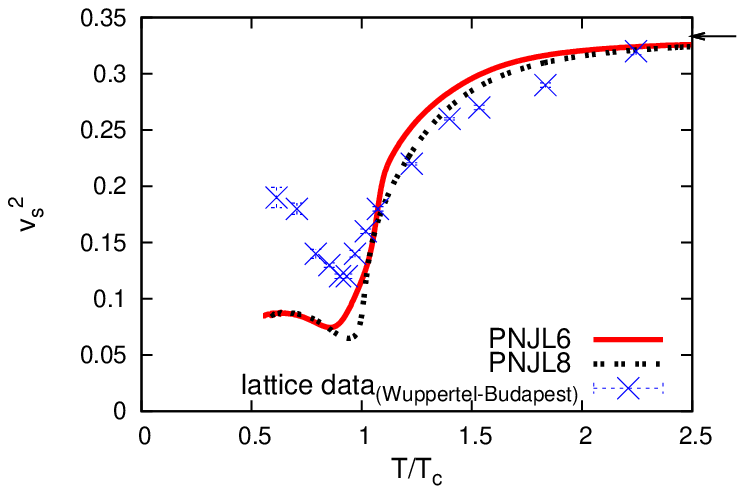}
\includegraphics[scale=0.58]{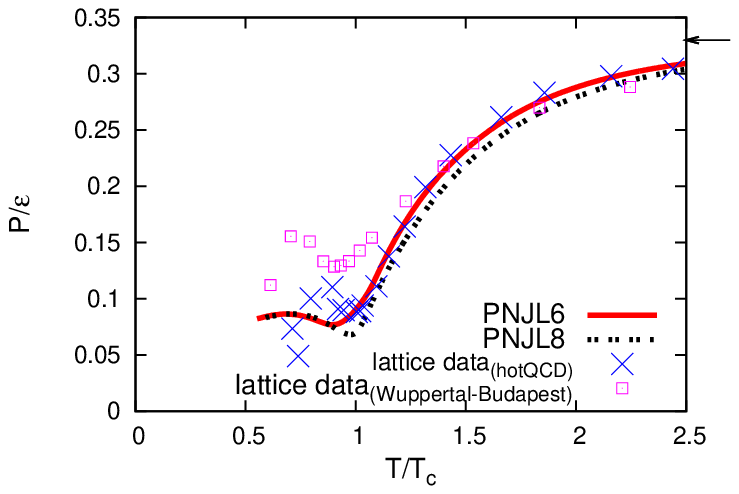}
\caption { Variation of $m1$, $m2$ and $m3$ with $T/T_C$, for 6q and 8q case.
The lattice data taken from Ref. \cite{gupta}. $C_V/T^3$, $v_s^2$ and 
$p/{\epsilon}$ as a function of $T/T_C$ are also shown. The lattice data 
is given by \cite{cheng1,fodor10}.} 
\label{moments}
\end{figure}
In figure (\ref{qns}) we show the variation of $c_2$, $c_4$ 
with $T/T_C$ for $\mu_q$ for both models and lattice data.
It can be seen that QNS ($c_2^q$) shows an order parameter like behavior. 
The high temperature value reaches almost 99\% of the ideal gas value.
The fourth order derivative $c_4^q$ shows a peak near the transition 
temperature $T_C$. 
The $c_4^S$ has a similar behavior as the $c_4^q$. However the peak appears
at much higher $T$ than $T_C$ and coincide with the temperature at which
$d\sigma_s /dT$ of heavy strange quark shows a maximum. 
The charge susceptibilities show similar behavior as quark number susceptibility
\cite{deb3}. We have plotted the kurtosis $\it{i.e.}$ the
ratio of $c_4^X/c_2^X$ for both type of potential (where $X=q, Q ~ \rm{or} ~ S$)
and compared with the lattice data. The plot for $c_4^q/c_2^q$ for 8q 
interaction shows more fluctuation near $T_C$ than 6q interaction.
For the strangeness fluctuation, first peak occurs at chiral transition for 
light flavors and second peak occurs when chiral transition occurs in strange
sector. At intermediate temperatures PNJL model overestimates the ratio
than LQCD result.
Since the volume of the fireball is unknown, we have constructed the ratios of
these cumulants, so that the volume is cancelled out.
We now plot the moments for both 6q and 8q interactions. Our results show 
monotonous behavior in the plots. However the lattice result show nonmonotonous behavior for certain range of chemical potential. This difference may be due to
the fact that our CEP is at much higher chemical potential region than the 
Lattice value \cite{gupta}.
The thermodynamic quantities like specific heat ($C_V$) and the
speed of sound ($v_s$) are shown in fig \ref{moments}. 
The softest point of the equation of state is found to be
$(p/\epsilon)_{min}\approx 0.07$ for PNJL6 and
$(p/\epsilon)_{min}\approx 0.06$ for PNJL8.
PNJL8 model gives better agreement with lattice data, which has its softest
point of equation of state as $(p/\epsilon)_{min}\approx 0.05$ \cite{cheng1}.
However the softest point in other lattice data comes out to be $\sim 0.13$ in 
\cite{fodor10}.
\vskip 0.05in
{\section{Conclusion}} 
\vskip -10pt
To conclude, we have studied the phase diagram and the fluctuations of 
different conserved charges of the PNJL and NJL model with eight-quark 
interaction. The eight-quark interaction term shifts the CEP to the 
low $\mu$ and high $T$ value. Furthermore our study concludes that the 
inclusion of eight-quark interaction is essential to limit $\mu_C/T_C$ 
below 2.5, as suggested by the recent lattice data. Fluctuations of different
conserved charges show significant behavior near the phase transition
temperature. 

\vskip 0.05in
{\section{Acknowledgement}}
\vskip -10pt
P.D. thanks CSIR for financial support.
A.B. thanks UGC (UPE and DRS) for support. The authors
thank Saumen Datta and Sourendu Gupta for useful discussions.

\end{document}